\documentclass[floatfix,aps,twocolumn,showpacs,preprintnumbers,amsmath,amssymb]{revtex4-1}

\usepackage[applemac]{inputenc}
\usepackage[T1]{fontenc}
\usepackage{amsfonts, amssymb, amsmath, dsfont}
\usepackage{graphicx}
\usepackage{lipsum}
\usepackage{float}
\usepackage{ulem}
\usepackage{color}

\newcommand{\bF}{\mathbf{F}}

\newcommand{\bE}{\mathbf{E}}

\newcommand{\bH}{\mathbf{H}}
\newcommand{\bcalE}{\boldsymbol{\mathcal{E}}}

\newcommand{\bcalH}{\boldsymbol{\mathcal{H}}}
\newcommand{\bS}{\mathbf{S}}
\newcommand{\bN}{\mathbf{N}}

\newcommand{\br}{\mathbf{r}}

\newcommand{\tk}{\tilde{k}}
\newcommand{\tq}{\tilde{q}}
\newcommand{\tn}{\tilde{n}}
\newcommand{\teps}{\tilde{\varepsilon}}

\newcommand{\lmax}{\ell_\mathrm{max}}
\def\d{\mathrm{d}}
\def\m{\mathrm{m}}
\def\eps{\varepsilon}
\def\Re{\mathrm{Re}}
\def\Im{\mathrm{Im}}

\begin{document}

%\title{Plasmonic force and torque on a spherical object}
\title{Transverse spinning of a sphere in a plasmonic field}

\author{Antoine Canaguier-Durand}
\affiliation{ISIS \& icFRC, University of Strasbourg and CNRS, 8 all\'{e}e Gaspard Monge, 67000 Strasbourg, France.}
\author{Cyriaque Genet}
\altaffiliation{Corresponding author: genet@unistra.fr}
\affiliation{ISIS \& icFRC, University of Strasbourg and CNRS, 8 all\'{e}e Gaspard Monge, 67000 Strasbourg, France.}

\begin{abstract}
We evaluate optical forces and torques induced by a surface plasmon to a sphere of arbitrary size, i.e. beyond the point-like dipolar limit.  Through a multipolar decomposition of the plasmonic field, we demonstrate that the induced torque is purely transverse to the plasmon propagation direction. Our approach removes the inherent ambiguities of the dipolar regime with respect to rotations and emphasizes the crucial role played by dissipation in the onset of the plasmonic torque. We also give realistic estimates of such plasmon-induced spinning of gold spheres immersed in water or air. 
\end{abstract}

\maketitle

\section*{Introduction}

The capacity to manipulate small objects through optical forces has impacted many research areas \cite{lewenstein2007ultracold,moffitt2008recent}. Light-induced spinning, in particular, has recently attracted a lot of attention \cite{padgett2011tweezers}, using circularly polarized and singular beams for rotating absorbing particles \cite{he1995direct,friese1996optical,o2002intrinsic,lehmuskero2013ultrafast} or unpolarized fields for chiral particles \cite{higurashi1994optically,higurashi1997optically,friese1998optical,eriksen2003spatial,liu2010light,liu2012chiral,canaguier2013mechanical}. In all such experiments, the induced angular velocity vector is along the beam propagation axis. But the importance of transverse rotations has been lately recognized as a way to induce localized shear stresses in fluids and thus to provide new tools and new probes for mechanobiology \cite{Parkin2007525}. While actual recent demonstrations of such rotations have involved rather complex ‘paddle-wheel' microstructures \cite{asavei2013optically} or potentially an optical ‘photonic-wheel' setup \cite{banzer2013photonic}, it has been emphasized theoretically that evanescent fields are interesting since they can readily spin transversely dissipative dielectric particles \cite{chang1998optical,canaguier2013force,bliokh2013extraordinary}.

In this context, surface plasmons (SP) are particularly appealing, given their strong potential for efficient optical manipulations in a great variety of configurations, ranging from SP assisted optical traps, propellors, sorters, etc. \cite{juan2011plasmon,wang2009propulsion,cuche2012brownian,cuche2013sorting,marago2013optical}. We have actually shown that due to the non-zero intrinsic spin expectation value of the plasmonic field, an SP mode can exert torques on dissipative dipoles in a transverse direction \cite{canaguier2013force}. In this article, we derive generalized forces and torques exerted by a plasmonic field on a sphere of arbitrary size. We show, through a multipolar expansion of the SP field and its reflection on the sphere, that the induced torque can genuinely spin the sphere, as soon as it is dissipative. With an angular velocity vector that turns out to be purely transverse with respect to the SP propagation direction, our scheme appears as a most simple alternative for the implementation of transverse rotations. Importantly, our multipolar approach removes the ambiguity of the meaning of rotation inherent to the dipolar regime, where the sphere is a point-like object. It appears then possible to evaluate angular velocities and frequencies of metallic spheres immersed in fluids, including therefore rotational Brownian motions. Such evaluations are relevant when aiming at observing such spinning effects. 

%\textcolor{red} {In this work, we will neglect higher-order scattering between the plane and the sphere, restricting ourselves to a first-order scattering approach. This approach, despite its limited range of validity (see more discussions below), is nevertheless commonly used as a most appropriate one to get qualitative and meaningful results. While multiple reflections between the plane and the sphere must be included if one aims for precise quantitative evaluations, we emphasize that our approach does lead to a clear physical picture of the dynamics at play in the plasmonic near field.}

\section{Plasmonic force and torque on a sphere}

\subsection{Surface plasmon spin}

\begin{figure}[htbp]
\centering%\vspace{4cm}
\includegraphics[width=0.48\textwidth]{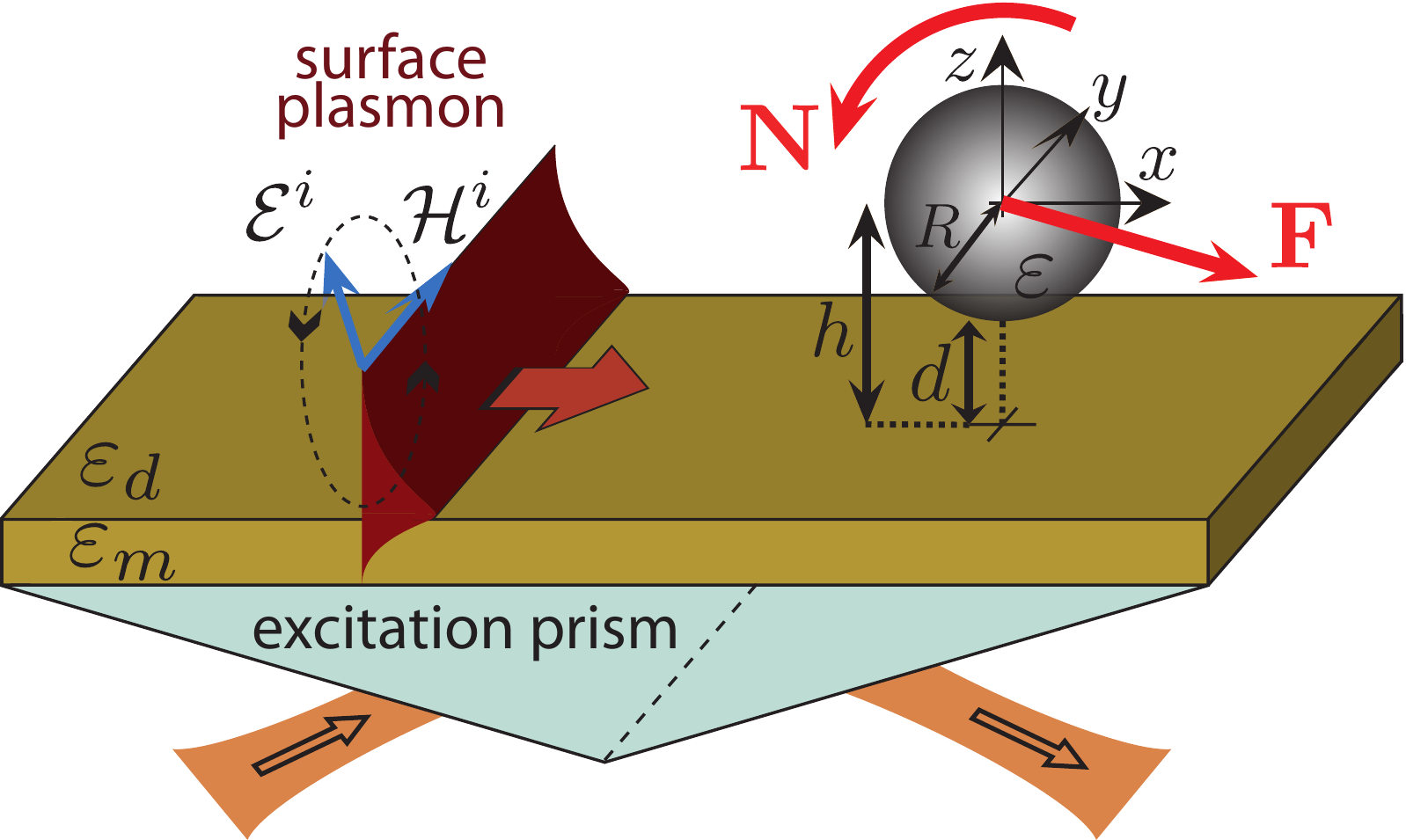} %
\caption{(Color online) Sketched configuration: an SP mode is launched in the $x$ direction at a metal / fluid (water) interface. The associated plasmonic field $\left( \bcalE^i, \bcalH^i \right)$ exerts a force $\mathbf{F}$ and a torque $\mathbf{N}$ on a sphere of radius $R$ immersed in the fluid at a distance $h$ measured from its center to the interface.}
\label{schema_situation}
\end{figure}

We consider an harmonic plasmonic field  \mbox{$\left( \bcalE^i (\br,t), \bcalH^i (\br,t) \right)= \Re \left( \bE^i(\br) e^{-\imath \omega t }, \bH^i(\br) e^{-\imath \omega t }  \right)$} with angular frequency $\omega$ and wavelength in vacuum $\lambda=\frac{2\pi c }{\omega}$. The SP mode is launched unidirectionally at an interface made of a metal, described by a complex permittivity $\eps_m$, and a non-dissipative fluid described by a real-valued dielectric function $\eps_d$. Practically, this can be done through total internal reflection of an incident laser beam –typically, using a prism in a Kretschmann-like configuration \cite{RaetherBook}, as sketched in Fig.~\ref{schema_situation}. A metallic sphere is then immersed in the fluid, localized slightly off the SP launching region and close enough to the metal so that it lies into the evanescent tail of the surface plasmon mode. This implies that the plasmonic field is freely propagating towards the sphere and incident on it  - see Fig.~\ref{schema_situation}. Note that this situation must be distinguished from surface plasmon mode coupling effects where a localized plasmon resonance pre-excited on the sphere couples to the SP mode at the planar metal film.

With the $x$ axis taken as the propagation direction, the SP field writes as:
\begin{align}\label{plasmonic_field2}
&\bH^i(x,y,z) = H^i   e^{\imath k x} e^{\imath q (z+h)}  \left(  0, 1 , 0 \right)^t \nonumber \\
&\bE^i(x,y,z) =  E^i  e^{\imath kx} e^{\imath q (z+h)}   \left( \tq , 0 , -\tk \right)^t  
\end{align} 
where $H^i=\sqrt{\eps_\d \eps_0/\mu_0}E^i$, and $h$ is the distance between the interface and the center of the sphere (Fig.~\ref{schema_situation}). The complex wavenumbers $k$ and $q$ of the plasmonic field
\begin{align*}
&k=k'+\imath k'' = \frac{\omega n_\d}{c} \sqrt{\frac{\eps_\m}{\eps_\d + \eps_\m}}  & \tilde{k} = \tk' + \imath \tk'' =\frac{kc}{n_\d \omega}  \\
&q=q' + \imath q'' = \frac{\omega n_\d}{c} \sqrt{\frac{\eps_\d}{\eps_\d + \eps_\m}}  & \tilde{q} = \tq' + \imath \tq'' = \frac{qc}{n_\d \omega} 
\end{align*} 
are determined by the frequency-dependent dielectric functions $\eps_\m,\eps_\d=n_\d^2$, $n_d$ being the fluid refractive index, and using for all numerical applications the optical data from \cite{palik1985handbook}. Let us recall that in order to have a well-defined propagating mode, the propagation length of the SP mode must be greater than its wavelength. This condition is fulfilled when $|\Re(k)|\gtrsim |\Im(k)|$, i.e. $|\Re(\eps_\m)| \gtrsim \eps_\d$. For a gold-water interface, this imposes $\lambda \gtrsim 480$ nm for the incident wavelength. In contrast, this condition is always fulfilled in the visible for an aluminium-water interface.

\begin{figure}[htbp]
\centering
\includegraphics[width=0.42\textwidth]{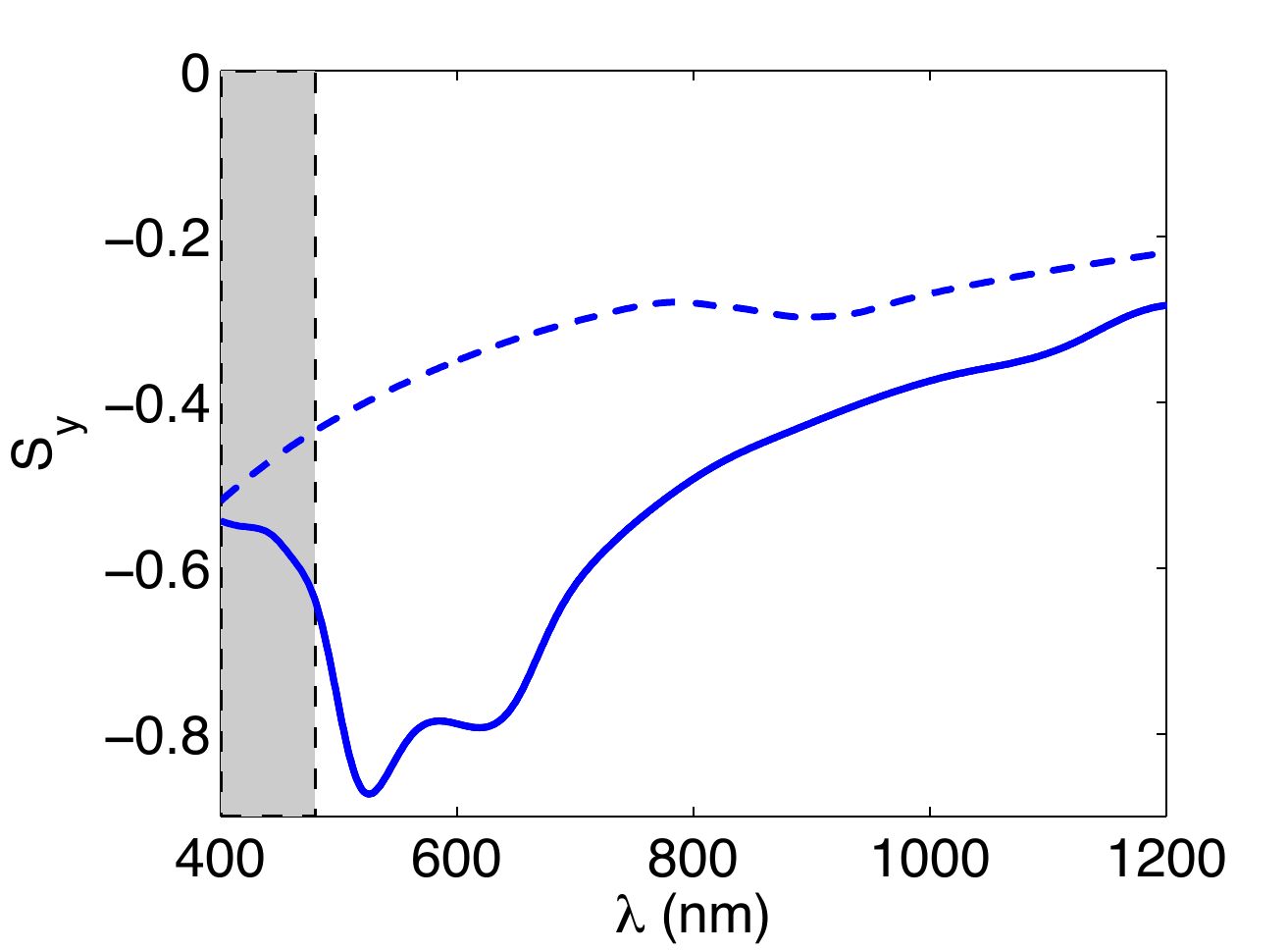} %
\caption{(Color online) Amplitude of the $y$-component $S_y$ of the spin expectation value $\mathbf{S}$ as a function of the incident wavelength $\lambda$, for a metal-water interface. The metal is gold (solid line) or aluminium (dashed-curve). The plasmon becomes ill-defined for $\lambda \lesssim 480$ nm (grey area) at a gold-water interface.}
\label{Sy}
\end{figure}

An interesting property of such transverse magnetic evanescent SP field is that the real-valued electric field $\bcalE$ rotates in the $(x,z)$ plane, as sketched in Fig.~\ref{schema_situation}. The SP field described by Eq.~(\ref{plasmonic_field2}) thus possesses a non-zero ellipticity in the $y$ direction \cite{bliokh2012transverse,canaguier2013force} and, as a consequence, has a spin expectation value
\begin{align}\label{S_y}
\mathbf{S} = -2 \frac{\tk'\tq'' - \tq' \tk''}{|\tk|^2 + |\tq|^2} \left( 0, 1, 0 \right)^t 
\end{align}
purely transverse along the $y$ axis. The $y$-component of $\mathbf{S}$ is evaluated in Fig.~\ref{Sy} as a function of the  wavelength $\lambda$, in the visible and close infra-red ranges, for gold- and aluminium-water interfaces. 
For both, this quantity is always negative, larger in magnitude for gold than for aluminium. In the former case, local maxima for $|S_y|$ are found at $\lambda\simeq$ 525 and 620 nm, while for higher wavelengths, $|S_y|$ slowly decreases. This parameter, an important property of plasmonic fields, has direct mechanical consequences. As shown in the dipolar regime, it is at the origin of the mechanical torque induced by an SP mode. We show below that the same holds for dissipative sphere of arbitrary sizes.

\subsection{Consequences for a Rayleigh particle}

We first recall the main results derived in \cite{canaguier2013force} related to force and torque exerted on a small sphere of radius $(R\ll \lambda)$ and complex dielectric function $\eps(\omega)$, immersed in the fluid at a distance $h$. In this Rayleigh regime, the sphere can be considered as a point-like dipole, whose polarizability $\alpha_0$ is described by the Clausius-Mossotti formula:
\begin{align}\label{alpha_0}
\alpha_0 = 4\pi\eps_0 R^3 \frac{\eps - \eps_\d}{\eps + 2 \eps_\d} 
\end{align}
where $\eps$ and $\eps_\d$ are the dielectric functions for the sphere and the fluid, respectively. The plasmonic force on an electric dipole can be written as a sum of reactive and dissipative components:
\begin{align}\label{Fdip}
\frac{\bF^\mathrm{dip}}{I_0} = & \frac{n_\mathrm{d}}{c} e^{ -2q'' h} \left(1+ 2(\tk'')^2 + 2(\tq'')^2\right) \nonumber \\
& \times  \left[  - \frac{\Re[\alpha_0]}{\eps_0}  \left(\begin{array}{c} k'' \\ 0  \\ q'' \end{array} \right) + \frac{\Im[\alpha_0]}{\eps_0}  \left(\begin{array}{c} k' \\ 0  \\ q' \end{array} \right)\right]
\end{align}
where $I_0=n_\mathrm{d}  \eps_0 c |E^i|^2/2$ is the incident field intensity in the fluid. These two components connect the real part of the sphere polarizability to the intensity gradient and the imaginary part to the phase gradient, respectively (see \cite{canaguier2013force} and references therein). The total force, in the $(x,y)$ plane, is directed along the propagation direction of the plasmon with an additional negative $z$-component, as sketched in Fig.~\ref{schema_situation} with a red arrow. The plasmonic torque
\begin{align} \label{Ndip}
\frac{\bN^\mathrm{dip}}{I_0} &= \frac{n_\d}{  c} e^{ -2q'' h} \left(1+ 2(\tk'')^2 + 2(\tq'')^2\right)\Im\left[\frac{\alpha_0}{\eps_0}\right] \mathbf{S}  
\end{align}
is proportional to both the dissipation in the dipole and the spin expectation value \cite{canaguier2013force,bliokh2013extraordinary}. It thus follows directly from Eq.~(\ref{S_y}) that the torque $\bN^\mathrm{dip}$ is purely transverse, directed along $(y<0)$ axis, as sketched in Fig.~\ref{schema_situation}.

Although the dipolar approach provides simple expressions that help capturing the specificities of the plasmonic force and torque, the physical interpretation of the results is ambiguous because, in the Rayleigh regime, the sphere is point-like and only characterized by its dipolar moment. It is thus unclear whether this torque induces a genuine mechanical spinning of the particle or only a spinning of its dipolar moment. In fact, the only consequence of the dipolar torque $\bN^\mathrm{dip}$ is the spinning of the sphere dipolar moment, in the same direction as the electric field but delayed from it \cite{canaguier2013force}. Moreover, the harmonic field assumption and the linear response description with a frequency-dependent polarizability $\alpha_0$ necessary lead to a harmonic dipolar moment that rotates, at all times, at the same angular frequency as the electric field. As a consequence too, an initial harmonic spinning is implicitly forced on the sphere dipolar moment, leaving no way to study a genuine torque induced on the sphere initially immobile. One major aim of this article is to remove such ambiguities with a multipolar treatment of the plasmonic scattering on the sphere. Only such a treatment gives the possibility to consider objects of finite sizes, initially at rest. 

\subsection{The Mie scattering on a sphere}

\begin{figure}[htb]
\centering
\vspace{0.4cm}
\includegraphics[width=0.4\textwidth]{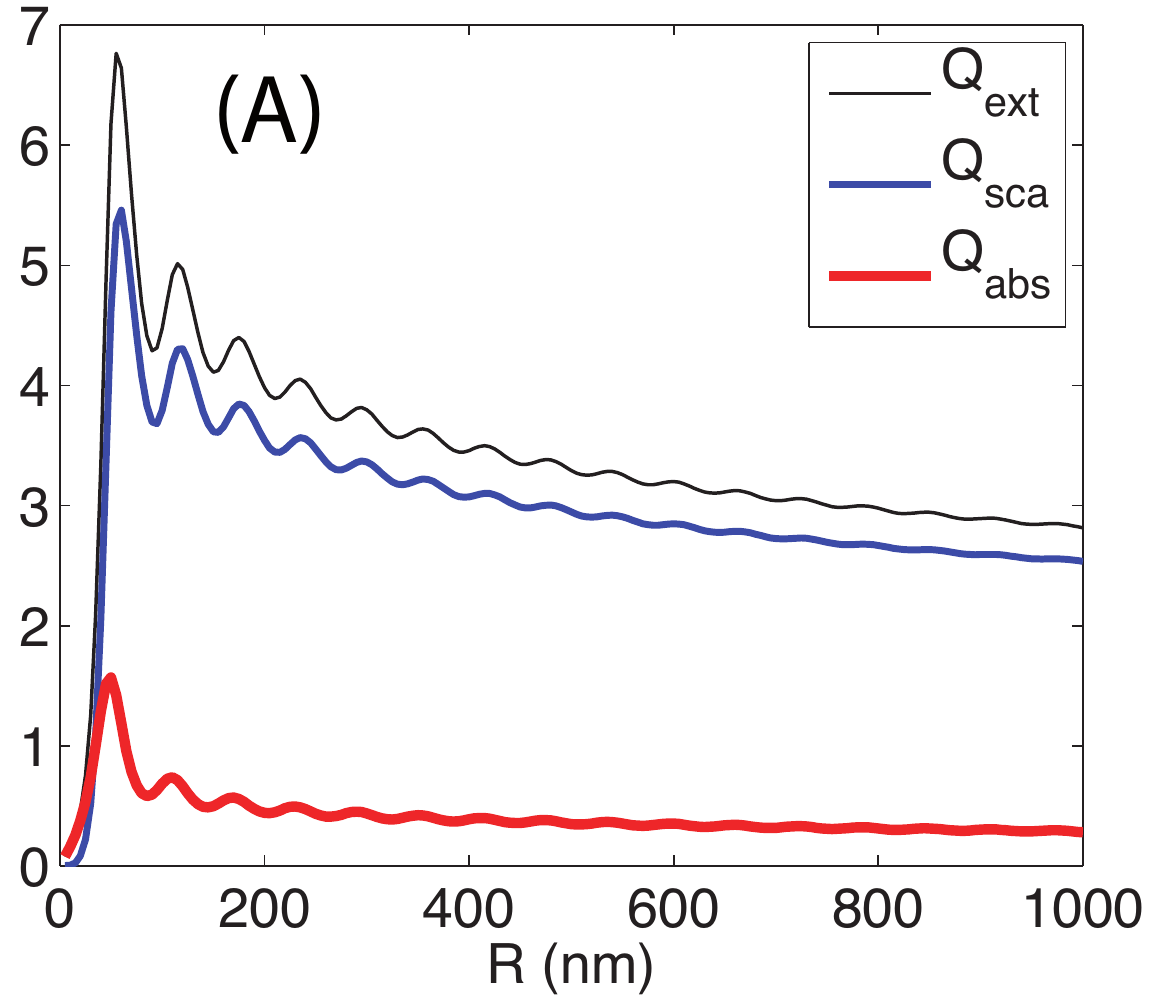} %
\includegraphics[width=0.4\textwidth]{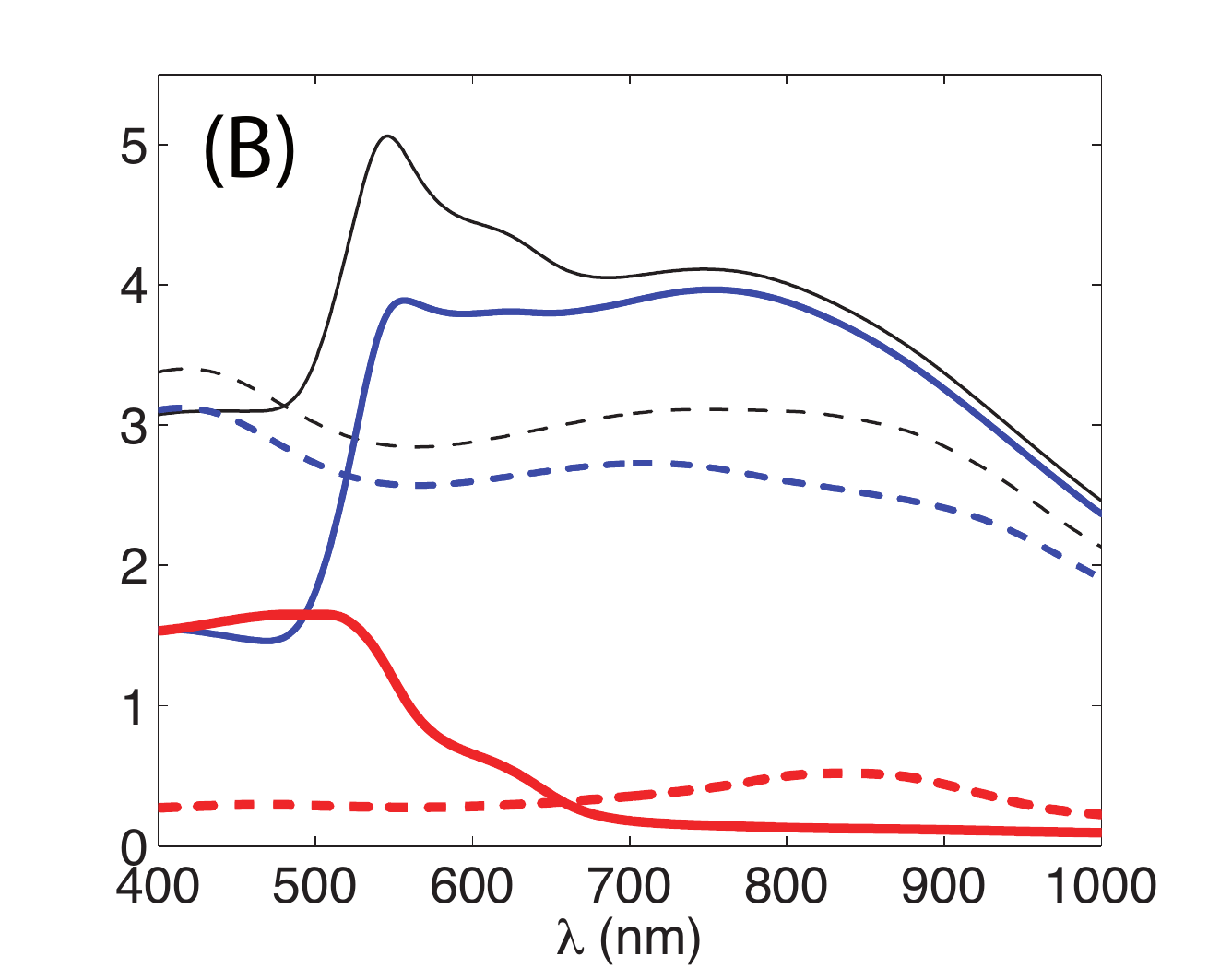} %
\caption{(A) Normalized $Q=C / (\pi R^2)$ extinction (thin dark curve), scattering (blue curve) and absorption (thick red curve) cross-sections for a gold sphere, as a function of its radius $R$. The incident wavelength is set to $\lambda=594$ nm. (B) Same quantities, as a function of the incident wavelength $\lambda$ for a 100 nm radius sphere made of gold (solid lines) or aluminium (dashed curves).}
\label{cross_sections}
\end{figure}

% Cross-sections
The multipolar nature of the scattering properties of a finite-size sphere is best illustrated through associated cross sections which characterize the energies removed from the incident light (extinction), radiated by the sphere (scattering) and absorbed inside the sphere (absorption). Because they do not depend on the amplitude and shape of the incident field, these quantities show the intrinsic optical response of the sphere. They can be obtained easily from the Mie coefficients of the sphere \cite{hulst1957light,bohren2008absorption} and are presented in Fig.~\ref{cross_sections} for a metallic sphere. 

When studying the dependence of the cross sections on the sphere radius (Fig.~\ref{cross_sections}(A)), small oscillations can be observed which are directly connected to the sequence of multipoles with increasing order. For an incident field of wavelength $\lambda=594$ nm, the first four maxima are approximately located at $R=$ 55, 115, 175 and 235 nm, displaying hence a pseudo-periodicity of about $\lambda / 10$.  

For gold and aluminium spheres of radius 100 nm that are typically used in recent experiments, and are falling beyond the Rayleigh regime in the visible range, the wavelength dependence of the different cross sections is shown in Fig.~\ref{cross_sections}(B). A key element for the analysis of the plasmonic torque is that the absorption cross section $C_\mathrm{abs}$ of a gold sphere (red solid curve) is maximum for $\lambda$ around 520 nm, with a shoulder around 620 nm and a rapid decrease above 700 nm. For aluminium,  $C_\mathrm{abs}$ is much smaller.  

%The imaginary part characterizes the energy dissipated by the sphere, and undergoes a sharp maximum for gold at wavelengths $\lambda \simeq 515$ nm, and a shoulder around 620 nm. For aluminium a small maximum is observed around 840 nm but is two orders of magnitude smaller. The dissipation peak yields two maxima for the real part of the polarizability for gold, at wavelengths $\lambda \simeq 540$ and 620 nm, whereas $\Re[\alpha_0]$ remains almost constant for aluminium in the visible range. 

\subsection{Optical force and torque applied to the sphere}

To go beyond the Rayleigh regime, a multipolar calculation is performed through the scattering formalism where the total electromagnetic field $\left( \bcalE, \bcalH \right)$ outside the sphere is taken as the superposition of a field $\left( \bcalE^i, \bcalH^i \right)$ incident on the sphere, and a scattered field $\left( \bcalE^s, \bcalH^s \right)$ going away from it. This multipolar decomposition is well suited to compute the field scattered by a sphere $\left( \bcalE^s, \bcalH^s \right)$ through the Mie coefficients given in Appendix~\ref{app:scattering}. The optical force and torque exerted on a sphere assumed to be at rest, are then obtained by integrating the Maxwell stress-tensor evaluated for this total field over the surface of the sphere.  

In this framework, the sphere appears as a local probe of the near field which perturbs the incoming SP field through the scattered field.  In return, the perturbed, i.e. total field acts on the sphere through optical force and torque. This amounts considering the perturbation of the incoming plasmonic field at the first order, neglecting the influence of multiple reflections induced between the sphere and the metal film. While multiple reflections between the plane and the sphere must be included if one aims for precise quantitative evaluations, this first-order approach is commonly used to get qualitative and meaningful results, as higher order corrections are mathematically and computationally demanding and only have a limited influence \cite{vsiler2011parametric}. Indeed, as shown in the case of total internal reflection on dielectric spheres, the effect of multiple reflections are only seen close to the particle Mie resonances, contributing only to a small fraction ($\sim 10\%$) of the force, and are totally negligible off-resonance \cite{song2001forces}. It is not straightforward to anticipate how the plasmonic force on a metallic particle would be affected but recent simulations have clearly shown that the effects of multiple reflections on the dynamics are only felt at very short distances and essentially affect the vertical component of the force \cite{wang2009propulsion,gaugiran2007spatial}.

%{\it Optical force applied to the sphere -}
The time-averaged force acting on the sphere can be obtained by integrating the normal component of the Maxwell stress tensor, relative to the total electromagnetic field, over the surface of the sphere \cite{barton1989theoretical,almaas1995radiation}. Its different components read, for an arbitrary incident field \footnote{This formula has been derived in \cite{barton1989theoretical}, and in \cite{almaas1995radiation,brevik2003radiation} with a misprint in the $z$-component.}:
\begin{widetext}
\begin{align}\label{Fxy_mie}
\frac{F_x + \imath F_y}{I_0} = &\frac{\imath \beta^2 R^2  n_\mathrm{d}}{2c} \sum_{\ell=1}^{\infty} \sum_{m=-\ell}^{\ell}  \left\{ \vphantom{\sqrt{\frac{(\ell+m+2)(\ell+m+1)}{(2\ell+1)(2\ell+3)}}} \right. \nonumber \\
& \sqrt{\frac{(\ell+m+2)(\ell+m+1)}{(2\ell+1)(2\ell+3)}} \ell (\ell+2)  \nonumber \\
& ~ ~ ~ ~ ~ ~ \times \left[  \left( 2 a_\ell a_{\ell+1}^{*} - a_\ell - a_{\ell+1}^{*} \right) A_{\ell,m} A_{\ell+1,m+1}^{*}  + \left( 2 b_\ell b_{\ell+1}^{*} - b_\ell - b_{\ell+1}^{*} \right) B_{\ell,m} B_{\ell+1,m+1}^{*} \right] \nonumber \\
&+\sqrt{\frac{(\ell-m+2)(\ell-m+1)}{(2\ell+1)(2\ell+3)}} \ell (\ell+2) \nonumber \\
& ~ ~ ~  ~ ~ ~ \times \left[  \left( 2 a_{\ell+1} a_{\ell}^{*} - a_{\ell+1} - a_{\ell}^{*} \right) A_{\ell+1,m-1} A_{\ell,m}^{*}  + \left( 2 b_{\ell+1} b_{\ell}^{*} - b_{\ell+1} - b_{\ell}^{*} \right) B_{\ell+1,m-1} B_{\ell,m}^{*} \right] \nonumber \\
& + \sqrt{(\ell+m+1)(\ell-m)} \nonumber \\
& ~ ~ ~ ~ ~ ~ \times \left[  \left( 2 a_{\ell} b_{\ell}^{*} - a_{\ell} - b_{\ell}^{*} \right) A_{\ell,m} B_{\ell,m+1}^{*}  - \left( 2 b_{\ell} a_{\ell}^{*} - b_{\ell} - a_{\ell}^{*} \right) B_{\ell,m} A_{\ell,m+1}^{*} \right]  \left. \vphantom{\sqrt{\frac{(\ell+m+2)(\ell+m+1)}{(2\ell+1)(2\ell+3)}}} \right\} 
\end{align}
\begin{align}\label{Fz_mie}
\frac{F_z}{I_0} =& - \frac{\beta^2 R^2 n_\mathrm{d}}{c} \sum_{\ell=1}^{\infty} \sum_{m=-\ell}^{\ell} \left\{ \vphantom{\sqrt{\frac{(\ell+m+2)(\ell+m+1)}{(2\ell+1)(2\ell+3)}}} \sqrt{\frac{(\ell-m+1)(\ell+m+1)}{(2\ell+1)(2\ell+3)}} \ell (\ell+2) \right. \nonumber \\
& ~ ~ ~ ~ ~ ~ \times \Im \left[  \left( 2 a_{\ell+1} a_{\ell}^{*} - a_{\ell+1} - a_{\ell}^{*} \right) A_{\ell+1,m} A_{\ell,m}^{*}  + \left( 2 b_{\ell+1} b_{\ell}^{*} - b_{\ell+1} - b_{\ell}^{*} \right) B_{\ell+1,m} B_{\ell,m}^{*} \right] \nonumber \\
& + m \Im \left[ \left( 2a_\ell b_\ell^{*} - a_\ell - b_\ell^{*}\right) A_{\ell,m} B_{\ell,m}^{*} \right]  \left. \vphantom{\sqrt{\frac{(\ell+m+2)(\ell+m+1)}{(2\ell+1)(2\ell+3)}}} \right\} 
\end{align} 
\end{widetext}
where $A_{\ell,m}$ and $B_{\ell,m}$ are the multipolar coefficients of the incident field, $a_\ell$ and $b_\ell$ are the Mie coefficients, and $\beta=n_\mathrm{d} \omega R /c$ an adimensional size parameter. 
%
%Let us emphasize that in the case of a surface plasmon as described in Eq.~(\ref{plasmonic_field2}), the $y$-component of the optical force has to be zero because of symmetry. 

%\subsection{Optical torque applied to the sphere}

%{\it Optical torque applied to the sphere -}
The expressions for the time-averaged torque applied to the sphere by the electromagnetic field can be found in \cite{barton1989theoretical}. In our case of a non-dissipative surrounding dielectric medium (real $n_\d$), this result can be simplified into:
\begin{widetext}
\begin{align}\label{Nxy_mie}
\frac{N_x + \imath N_y}{I_0} = & - \frac{\beta R^3  n_\mathrm{d}}{2c} \sum_{\ell=1}^{\infty} \sum_{m=-\ell}^{\ell} \ell (\ell+1) \sqrt{(\ell-m)(\ell+m+1)} \nonumber \\
& ~ ~ ~ ~ ~ ~ \times \left[  \left( 2 a_\ell a_{\ell}^{*} - a_\ell - a_{\ell}^{*} \right) A_{\ell,m} A_{\ell,m+1}^{*}  + \left( 2 b_\ell b_{\ell}^{*} - b_\ell - b_{\ell}^{*} \right) B_{\ell,m} B_{\ell,m+1}^{*} \right]  
\end{align}
\begin{align}\label{Nz_mie}
\frac{N_z}{I_0} = & - \frac{ \beta R^3  n_\mathrm{d}}{2c} \sum_{\ell=1}^{\infty} \sum_{m=-\ell}^{\ell} m (\ell+1)   \nonumber \\
& ~ ~ ~ ~ ~ ~ \times \left[  \left( 2 a_\ell a_{\ell}^{*} - a_\ell - a_{\ell}^{*} \right) A_{\ell,m} A_{\ell,m}^{*}  + \left( 2 b_\ell b_{\ell}^{*} - b_\ell - b_{\ell}^{*} \right) B_{\ell,m} B_{\ell,m}^{*} \right]  ~ .
\end{align}
\end{widetext}

\section{Numerical evaluations}

These expressions for the force -Eq.~(\ref{Fxy_mie},\ref{Fz_mie})- and the torque -Eq.~(\ref{Nxy_mie},\ref{Nz_mie})- are general and we now consider the specific case of an incident plasmonic field. In order to match with the symmetry of the sphere, the plasmonic field can be expanded in spherical modes through multipolar coefficients $A_{\ell,m}, B_{\ell,m}$ (see Appendix~\ref{app:multipolar} and references therein). 

The force and torque expressions derived above are exact but a truncation of the series to a finite number of modes is necessary for numerical evaluations. In practice, the relative size of the sphere compared to the incident wavelength dictates how much modes must be included to have an accurate estimation of the infinite series: the larger the ratio $R/\lambda$, the slower the series converges, and thus the larger cut-off $\lmax$ for the sum over $\ell$ one needs to choose to evaluate Eq.~(\ref{Fxy_mie},\ref{Fz_mie}) and Eq.~(\ref{Nxy_mie},\ref{Nz_mie}). The results obtained in this paper used a cut-off up to $\lmax=30$, sufficient to deal with spheres of radii $R \lesssim 1~\mu$m in the visible range. Larger spheres ($R \lesssim 20 ~\mu$m) could be treated with a cut-off more or less proportionally larger, and up to $\lmax\sim 500$ with a reasonable numerical cost \cite{canaguier2010thermal,canaguier2011developpement}.

We also note that all the plasmonic forces and torques turn directly proportional to the incident intensity $I_0=n_\d \eps_0 c |E^i|^2 / 2$. We therefore introduce for simplicity the reduced force $\mathbf{f}=\mathbf{F}/I_0$, and the reduced torque $\mathbf{n}=\mathbf{N}/I_0$ that will be used hereinafter. Additionally, in this first-order scattering approach, both the force and the torque turn out to scale with the exponential decay factor $e^{-q''d}$ through the incident field coefficients $A_{\ell,m}, B_{\ell,m}$ given  in appendix in Eq.~(\ref{AlmBlm}). This factor is determined on the plane-sphere separation distance $d$, which implies that force and torque can be normalized by this factor, the obtained ratio becoming constant for all distances. In the following of the section, we will therefore evaluate the force and the torque fixing this scaling factor to one for the sake of simplicity.

\subsection{Plasmonic force applied to a metallic sphere}

\begin{figure}[htbp]
\centering%\vspace{4cm}
\includegraphics[width=0.45\textwidth]{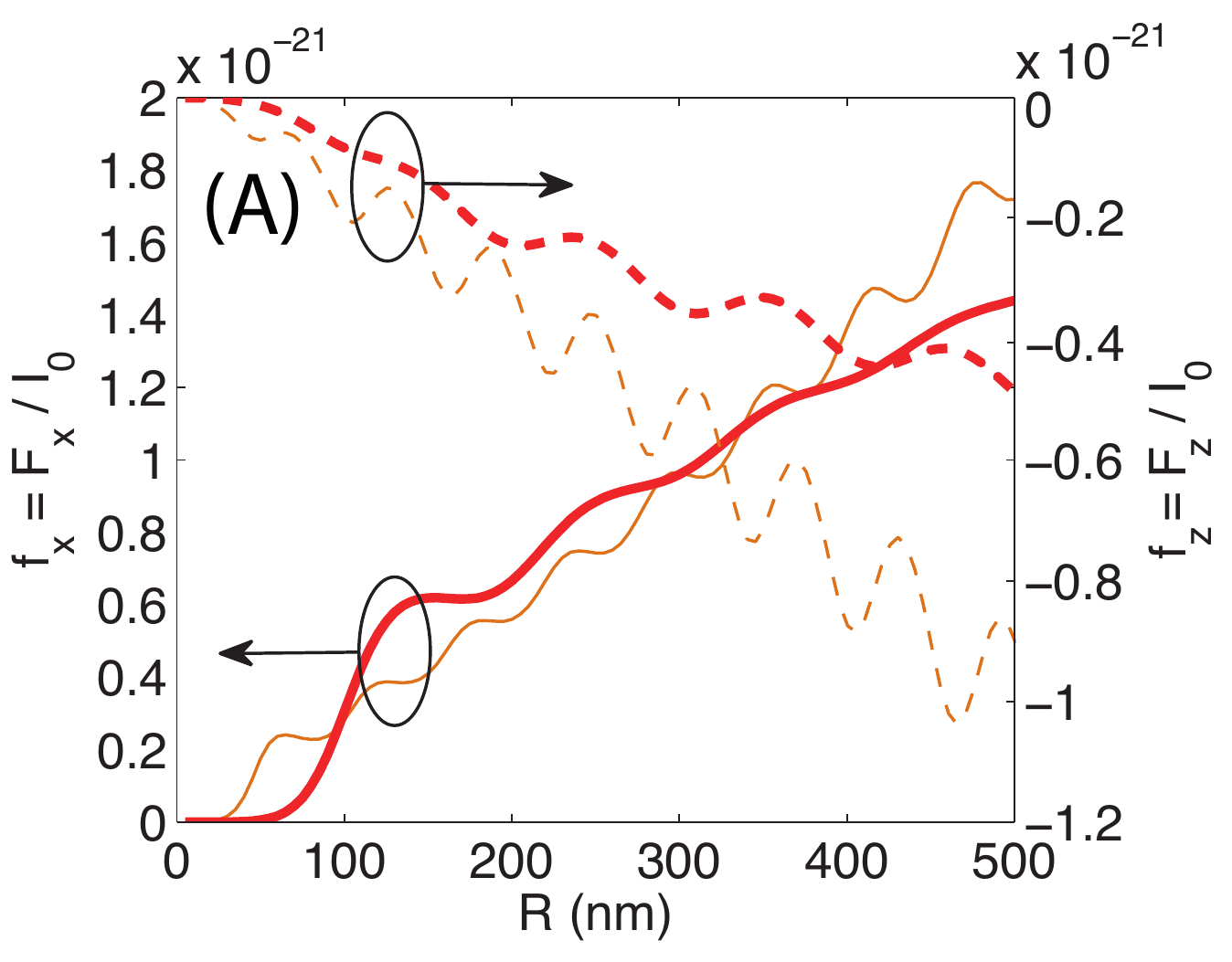} %
\includegraphics[width=0.38\textwidth]{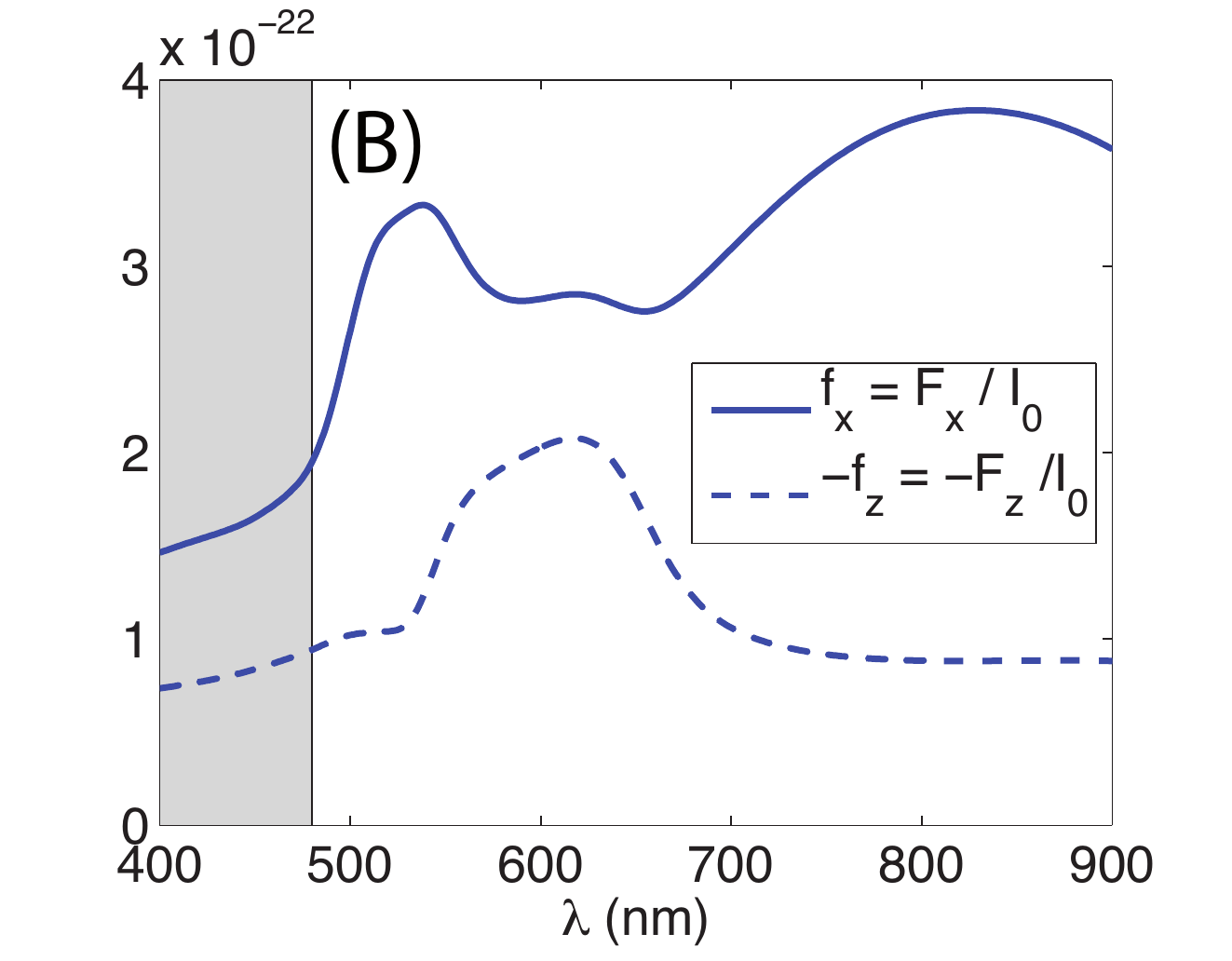}
\caption{(A) Normalized plasmonic force exerted on a gold sphere in the $x$ (left axis; solid lines) and $z$ (right axis; dashed-curves) directions, as a function of the sphere radius $R$. Two illumination wavelengths are considered, 594 nm (thin orange curves) and 980 nm (thick red curves). The plasmon is launched at a gold-water interface and $e^{-q''d}\equiv 1$. (B) Same quantities, but as a function of the incident wavelength $\lambda$, for a fixed sphere radius $R=100$ nm. The plasmon is ill-defined over the grey area.}
\label{F_594_980_vs_R}
\end{figure}

The results of the evaluations for the optical force in the $x$ and $z$ directions from Eqs.~(\ref{Fxy_mie}-\ref{Fz_mie}) on a gold sphere are presented in Fig.~\ref{F_594_980_vs_R}(A) as a function of the sphere radius $R$ (see also \cite{cuche2013sorting}). Here we fix the distance $h=R$ (see Fig.~\ref{schema_situation}) and we consider an incident plasmonic field launched at a gold-water interface with two incident wavelengths $\lambda=$ 594 and 980 nm. As expected from symmetry considerations, the force in the $y$ direction is zero in this case.  
For the others $x$ and $z$ components, the force globally increases in magnitude with oscillations every $\Delta R \simeq \lambda/10$. Such oscillations are typical of multipolar resonance effects as observed in Fig.~\ref{cross_sections}(A) and are thus due to the spherical geometry and the metallic nature of the object. These oscillations have been key when optically sorting metallic nanoparticles of different sizes, as we demonstrated experimentally \cite{cuche2013sorting}.

The dipolar limit sets a direct relation between the spin expectation value and the radiation pressure through the orbital part of the Poynting vector \cite{canaguier2013force}. Such a connection can still be seen in the multipolar regime, as observed in Fig.~\ref{F_594_980_vs_R}(B). There, with a sphere radius fixed at $R=100$ nm, the wavelength dependences of the amplitudes of the force components appear directly related to the spin expectation values displayed in Fig.~\ref{Sy} with local maxima of both $f_x$ and $|S_y|$ at 540 and 620 nm. For the vertical force component $f_z$, this influence is less critical, as the Poynting vector is almost horizontal for a plasmonic field: we only observe here an enhancement of the pulling down effect of the plasmonic field by a factor 2 between 550 and 650 nm.

\subsection{Plasmonic torque applied to a metallic sphere}

\begin{figure}[htbp]
\centering%\vspace{4cm}
\includegraphics[width=0.42\textwidth]{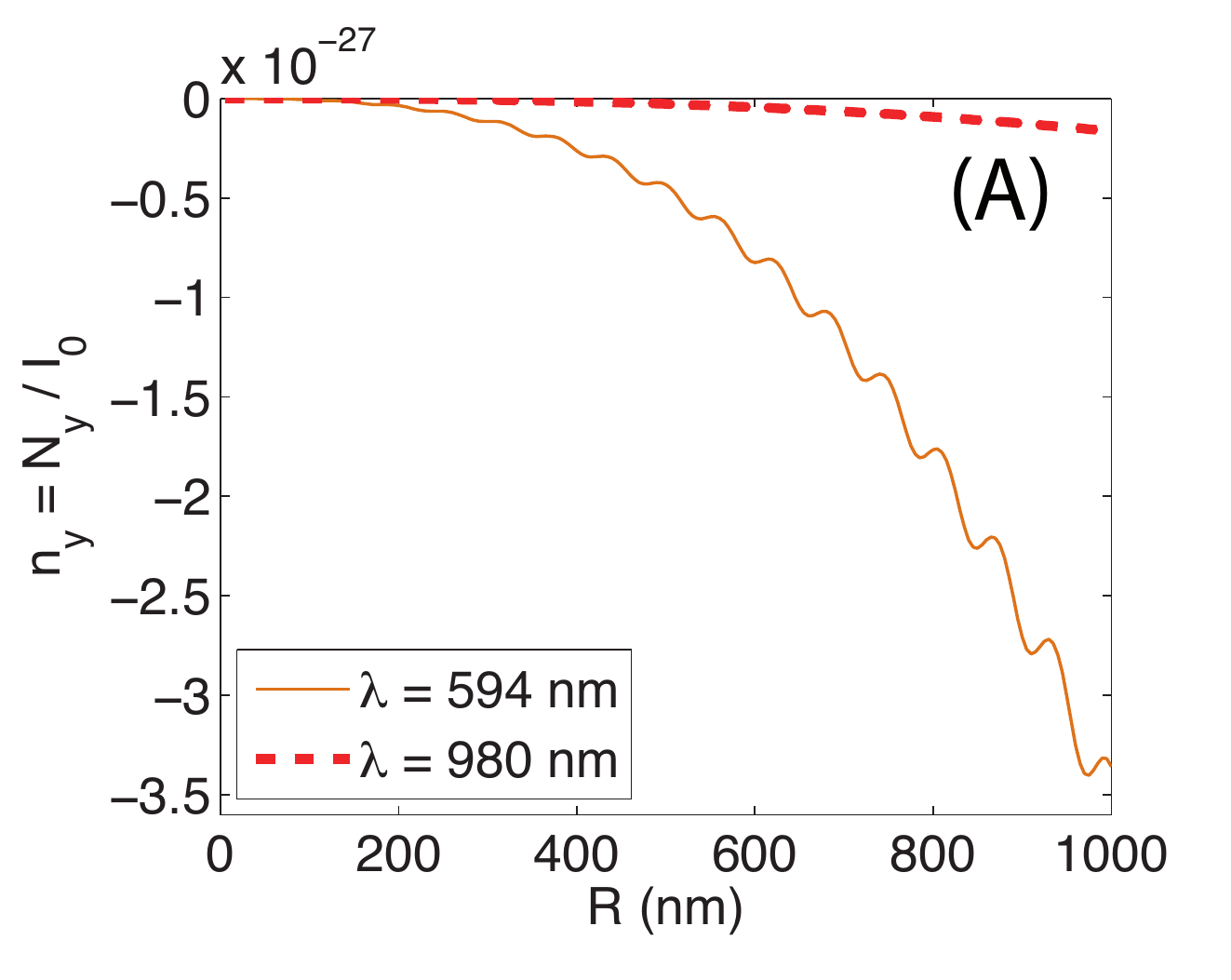} %
\hfill
\includegraphics[width=0.42\textwidth]{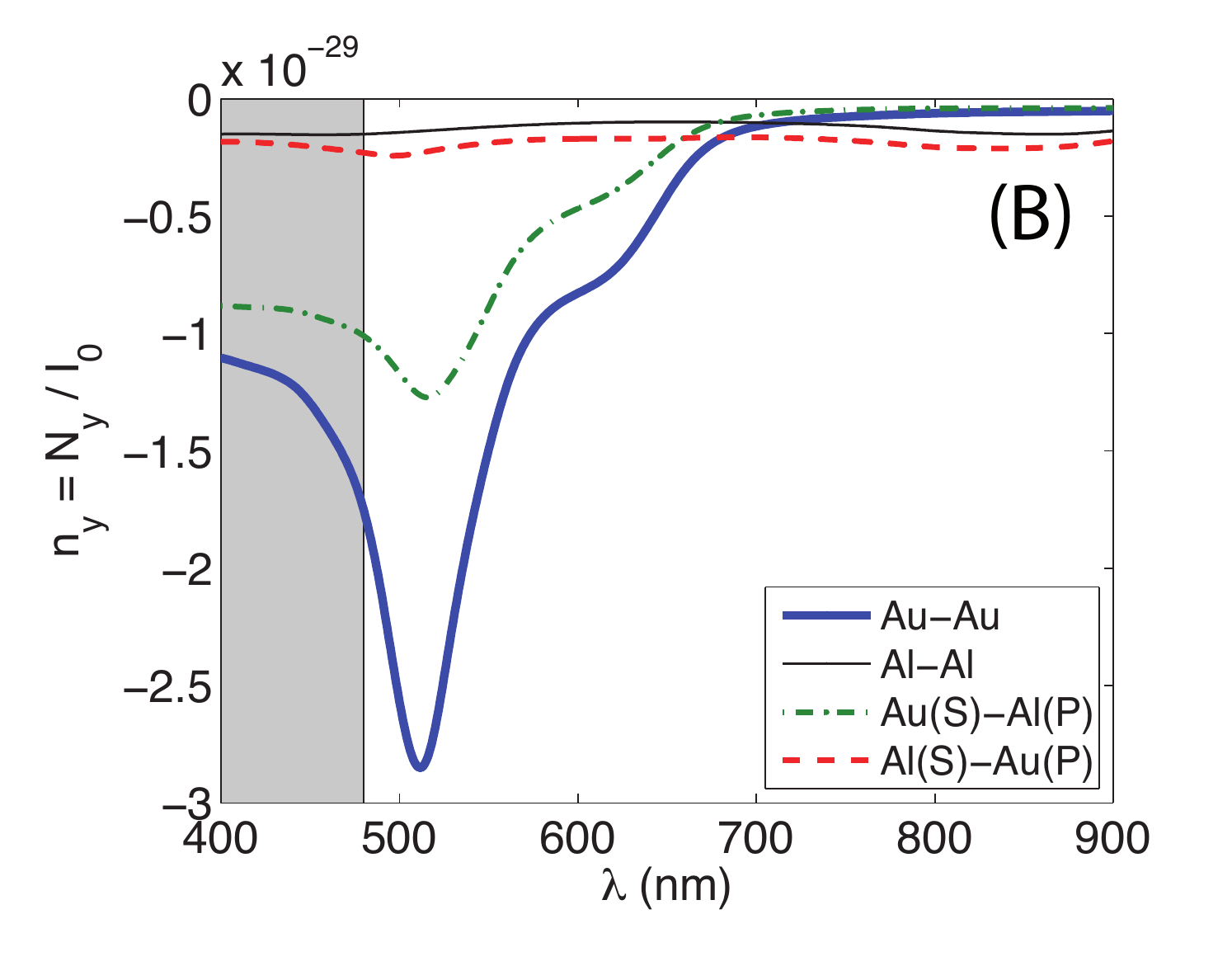}
\caption{(A) Normalized transverse plasmonic torque $n_y$ applied in the $y$ direction on a gold sphere, as a function of the sphere radius $R$. Two illumination wavelengths 594 and 980 nm are considered, and the plasmon is launched at a gold-water interface. (B) Same quantity, as a function of the incident wavelength $\lambda$, for a 100 nm sphere radius. Gold (Au) and aluminium (Al) are considered for both the sphere (S) and the metal plane (P). In both graphs the decay scaling factor $e^{-q''d}$ is set equal to one. The plasmon is ill-defined over the grey area.}
\label{Ny_594_980_vs_R}
\end{figure}

As the central result of this article, we evaluate the torque induced on a metallic sphere by a plasmonic field. Remarkably in this case, the $(x,z)$ plane symmetry constitutive of the plasmonic field yields that $N_x=N_z=0$ and thus leaves the sole possibility of a purely transverse torque in the $y<0$ direction. In the configuration of Fig.~\ref{schema_situation} (with the decay scaling factor $e^{-q''d}$ set equal to one), the normalized torque component $n_y$ is evaluated in Fig.~\ref{Ny_594_980_vs_R}(A) as a function of the sphere radius $R$, for two plasmon wavelengths 594 and 980 nm. As expected from the dipolar approach, the torque turns negative for all values of $R$ and $\lambda$, as drawn in Fig.~\ref{schema_situation}. It is crucial to note that the evaluation of a non-zero plasmonic torque in the multipolar approach removes the initial ambiguity brought forward by the dipolar limit. It shows indeed that the plasmonic field is genuinely able to spin mechanically small objects that are initially at rest. 

Like the force, and for the same reason, the torque displays small $\Delta R \simeq \lambda/10$ oscillations seen in Fig.~\ref{Ny_594_980_vs_R}(A).  The torque is stronger in magnitude at $\lambda=594$ nm than at $\lambda = 980$ nm, a point that is easily explained when looking at the wavelength dependence of the torque presented in Fig.~\ref{Ny_594_980_vs_R}(B) for a sphere of radius $R=100$ nm.  The torque indeed is maximal at $\lambda \simeq 520$ nm, with a shoulder around 620 nm, two features that can be traced back to two different sources: the spin expectation value of the plasmonic field presented in Fig.~\ref{Sy} and the absorption cross section of the sphere presented in Fig.~\ref{cross_sections}. Accordingly, the torque decreases for larger wavelengths, explaining the strong differences in strength between incident wavelengths of 594 nm and 980 nm.  

Because interband transitions play a crucial role in the generation of the ellipticity of the plasmonic field, and hence affect directly the associated spin expectation value, it is particularly interesting to check for metal dependence of the plasmonic torque. In this context, aluminium is an interesting metal \cite{ChimentoPhD}. The signatures of its small absorption losses at wavelength shorter than $600$ nm and its peculiar parallel-band transition at ca. $825$ nm can be seen in the cross-sections presented in Fig.~\ref{cross_sections}(B). Keeping identical configurations with $e^{-q''d}\equiv 1$, the torque is evaluated with aluminium for the sphere and/or the metal plane along which the SP mode is launched. The different curves show that the torque is strongest when both the sphere and the plate are made of gold, because in this case both the spin of the plasmonic field and the dissipation in the sphere are maximized. When the sphere is made of aluminum (red solid line and dark dashed curve), the torque is very small because the dissipation in the sphere is small. The mixed-case of a gold sphere and a aluminium-water plasmon (green curve) still demonstrate a substantial torque, as the spin of the plasmon shown in Fig.~\ref{Sy} is only twice smaller than the one obtained with a gold-water interface. As a consequence, we conclude that the plasmonic torque on a spherical object relies mainly on two ingredients: the spin expectation value $\bS$ of the incident plasmonic field and the dissipation in the sphere. This is precisely the conclusion that can be reached in the dipolar regime, but only qualitatively.

This study of finite size spheres allows an alternative interpretation of the plasmonic torque induced on a metallic sphere from the point of view of moving free-electrons. The SP mode, being an electronic density wave traveling in the $x$ direction, will induce and drive, through influence, an electronic density in the sphere in the $x$ direction as well. As a consequence of dissipation, these moving electrons are slowed down and transfer their momentum to the metal lattice. The bottom of the sphere is thus accelerated in the $x$ direction, which makes the sphere rotate around the $y$ axis. The negative sign obtained for the torque in the scattering evaluation is fully consistent with this view, making the torque transmitted by electrons equivalent to a non-contact gears mechanism.

\subsection{Plasmonic torque applied to a dielectric sphere}

\begin{figure}[htbp]
\centering%\vspace{4cm}
\includegraphics[width=0.42\textwidth]{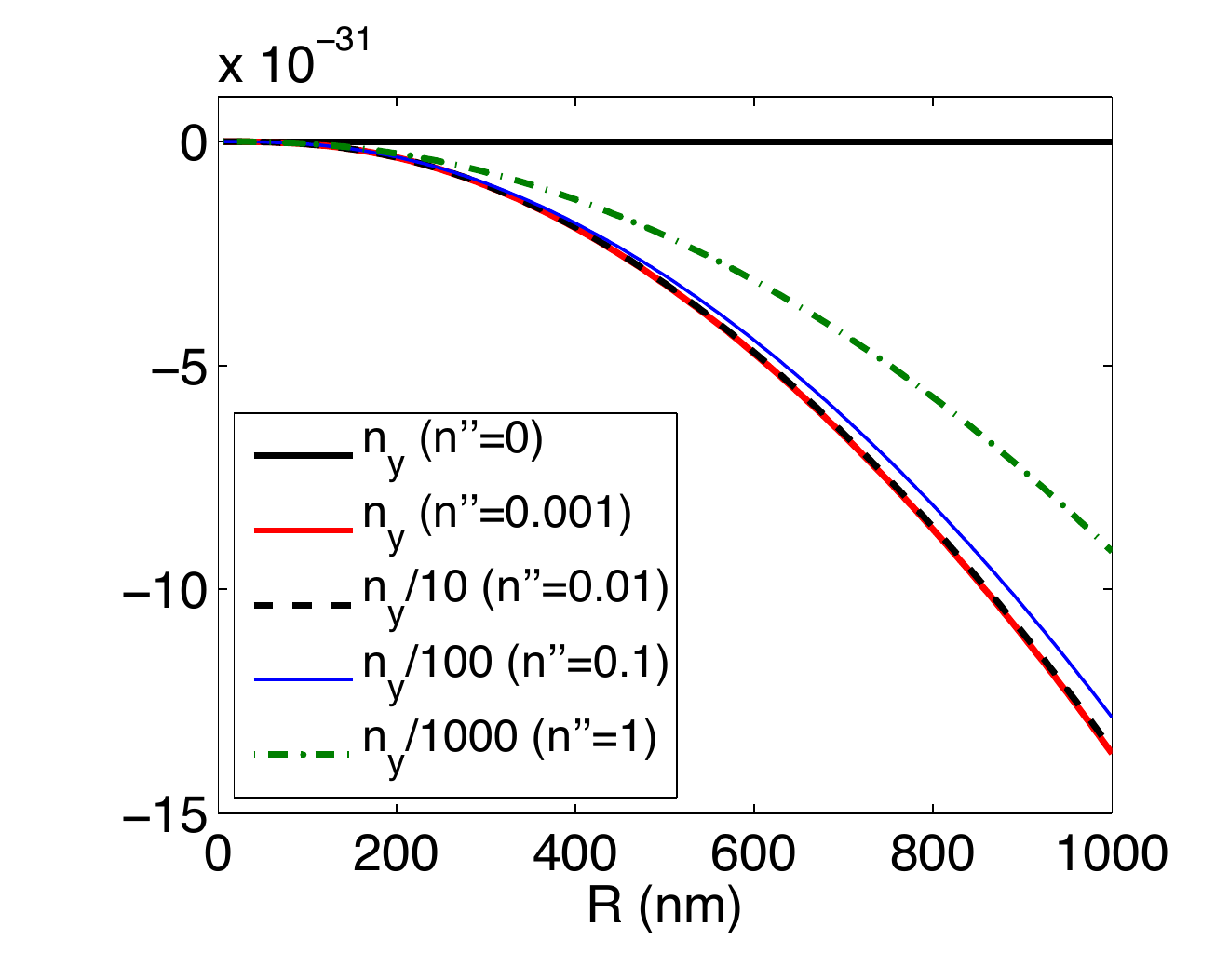} %
\caption{Normalized transverse plasmonic torque $n_y$ exerted on a dielectric sphere as a function of the sphere radius $R$. The incident wavelength is 594 nm and the refractive index of the sphere is $n=1.5+\imath n''$, with $n''=0$ (upper dark curve), 0.001 (lower red solid curve), 0.01 (lower dashed curve), 0.1 (thin blue curve), and 1 (dash-dot green curve). The torques for the three later cases are divided by 10, 100 and 1000 for clarity, respectively.}
\label{Ny_dielectric}
\end{figure}

This physical picture yields that dissipation in the material of the sphere is necessary for the torque to be induced. A simple way to test this idea is to see the effect of a progressive increase of the dissipation starting from a lossless dielectric sphere. We thus here evaluate the torque for dielectric spheres characterized by a complex refractive index $n=n'+\imath n''$. For $n''=0$, the sphere is non-dissipative and no torque should be observed. On the opposite, an increasing $n''>0$ should yield an increasing torque, up to the point where the reflectivity is too poor for the mechanical action to be efficient. This is fully verified by our numerical evaluations gathered in Fig.~\ref{Ny_dielectric} for increasing values of $n''$. 
Starting from the absence of torque for the non-dissipative sphere (solid dark curve), a slight increase to $n''=0.001$, a typical value for polystyrene spheres \cite{ma2003determination} yields a non-zero torque that smoothly increases with the radius $R$ of the sphere. Compared to the case of a metallic sphere, the applied torque is much weaker (more than three orders of magnitude) and the dependence of the torque on the radius $R$ does not undergo any rapid oscillations. When increasing the dissipation by successive orders of magnitudes, the torque is almost proportionally increased, as seen from the fact that the torques normalized by the corresponding factor are very close to the curve for $n''=0.001$. For the last two highly dissipative cases $n''=0.1$ and $n''=1$, the torque is shown to increase slower than the dissipation, indicating the beginning of a saturation effect for poor reflectors.

These results agree with the interpretation of a torque transferred from moving damped electrons. In the case of non-dissipative materials, the electrons do rotate around the center of the sphere but they do so freely and therefore do not lead to any torque. This is precisely what is to be understood from the dipolar analysis, where the point-like sphere has its electric dipole moment rotating in phase with the electric field, with no applied torque. 
In the dissipative case, an increasing damping of moving electrons leads to an increased transfer of momentum to the object and therefore an increased torque. For very dissipative materials, the electrons are damped before they acquire sufficient momentum, which could lead to the saturation effect observed in Fig.~\ref{Ny_dielectric} for high values of $n''$.

%\sout{Note that multiple reflections will only modify our predictions on a quantitative level, with a limited impact. Indeed, as shown in the case of total internal reflection on dielectric spheres, the effect of multiple reflections are only seen close to the particle Mie resonances, contributing only to a small fraction ($\sim 10\%$) of the force, and are totally negligible off-resonance\cite{song2001forces}. It is not straightforward  to anticipate how the plasmonic force on a metallic particle would be affected but recent simulations have clearly shown that the effects of multiple reflections on the dynamics are only felt at very short distances and essentially affect the vertical component of the force} \cite{wang2009propulsion,gaugiran2007spatial}.

\section{Plasmonic rotation of a sphere in a fluid}

It turns important to study the effect of the plasmonic torque on an immersed sphere since direct optical manipulation of particles is most efficient in fluids, essentially due to buoyancy. It is then crucial to account for a random Langevin torque $\bN_\mathrm{th}$, source of an inevitable rotational Brownian motion thermally induced by the surrounding fluid \cite{kalmykov1996rotational}. The dynamic equation for the instantaneous angular velocity vector $\mathbf{\Omega}$ of the sphere then writes
\begin{align}\label{dynamic_equation}
I \dot{\mathbf{\Omega}} + \gamma \mathbf{\Omega} = \bN + \bN_\mathrm{th}
 \end{align}
where $I=2MR^2/5$ is the inertia moment of the sphere of mass $M$ and $\gamma=8\pi\eta R^3$ is an estimation of the friction coefficient for a sphere of radius $R$ in a fluid of viscosity $\eta$. Here, we have removed all translational motions and forces by assuming that the center of mass of the sphere is at rest within the fluid. Remarkably on the stochastic equation (\ref{dynamic_equation}), the effects of the optical torque $\bN$ and the thermal torque $\bN_\mathrm{th}$ can be treated separately by introducing the deterministic (averaged) $\mathbf{\Omega}_\mathrm{d}=\langle \mathbf{\Omega} \rangle$ and stochastic $\mathbf{\Omega}_\mathrm{s}=\mathbf{\Omega} - \mathbf{\Omega}_\mathrm{d}$ parts of $\mathbf{\Omega}$ \cite{cuche2012brownian}. With this separation, it can be shown that the thermal torque $\bN_\mathrm{th}$ only affects the angular velocity vector $\mathbf{\Omega}$ by merely adding a Gaussian noise with a variance $\sigma=\frac{k_B T}{I}$ in every directions. 

The deterministic part is given by a constant torque $\bN$ directed in the $y<0$ direction. The plasmonic torque starts spinning the sphere up to an angular velocity limit $\Omega_\infty = N_y/\gamma$ reached within a time constant $\tau = I /\gamma$. Note that for sub-micron spheres immersed a viscous environment such as water corresponding to low Reynolds number, $\tau\sim 10^{-7}$ s meaning that friction prevails over inertia. In contrast, within tenuous environments, inertial effects have to be accounted for.   
\begin{figure}[htbp]
\centering%\vspace{4cm}
\includegraphics[width=0.42\textwidth]{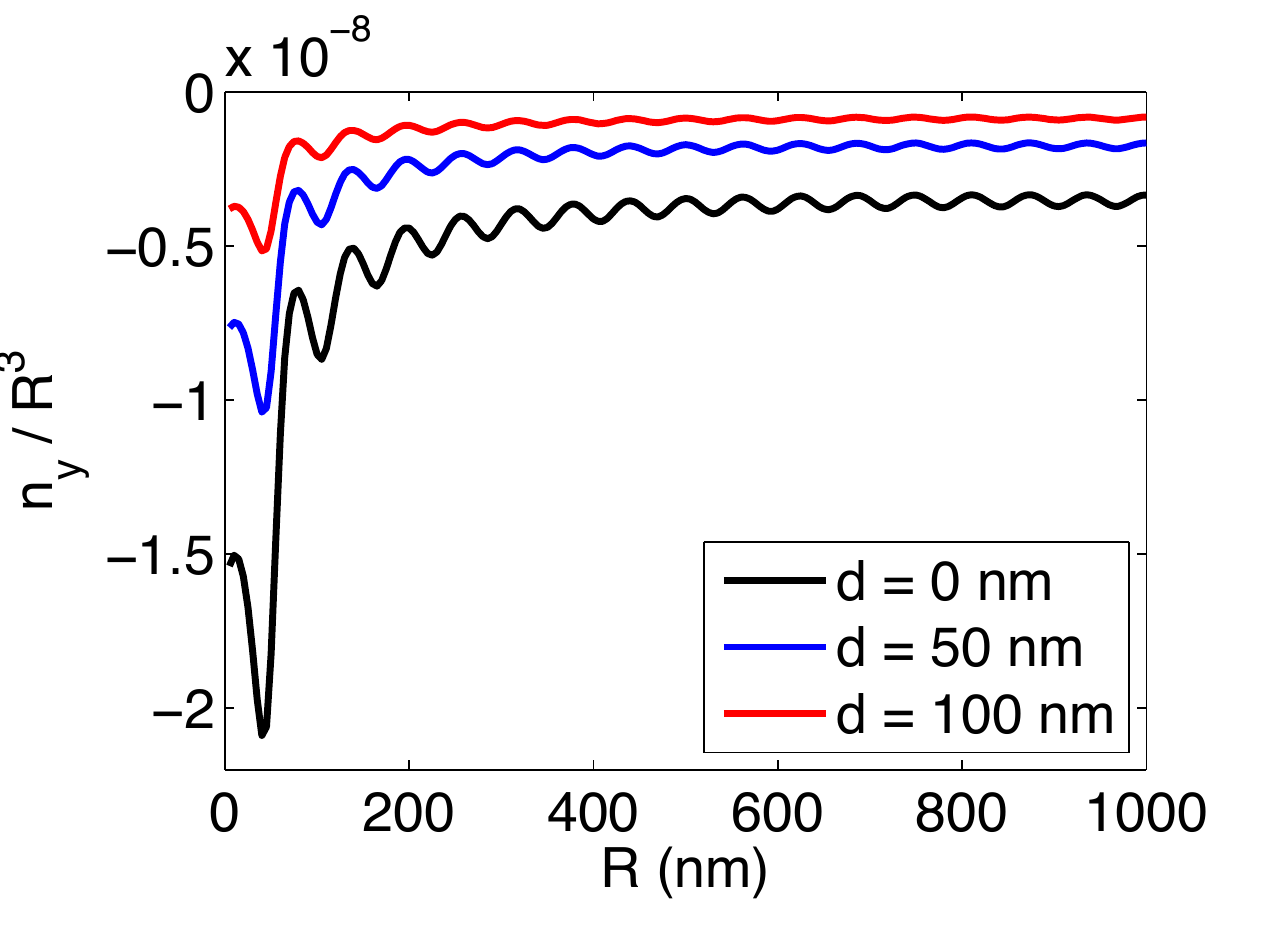} %
\caption{Normalized transverse torque $n_y$ divided by $R^3$, as a function of the radius $R$ of the sphere. Three separation distances considered are: $d =$ 0 (lower dark curve), 50 (blue curve), and 100 nm (upper red curve) between the gold plate and the bottom of the sphere.}
\label{omega_lim_gold}
\end{figure}

It is interesting to note that the normalized quantity $n_y/R^3$ scales precisely like the angular velocity limit $\Omega_\infty$. As shown in Fig.~\ref{omega_lim_gold} for three different separation distances $d=h-R$, $n_y/R^3$ globally converges to a constant for large metallic spheres, independently of $d$. As a consequence, for a given distance $d$ to the interface the torque $N_y$ scales as $R^3$ for large metallic spheres, and the spinning angular velocity $\Omega_\infty$ does not depend on the size of the sphere for $R\gtrsim 400$ nm. These results have been given within a first-order scattering model but their physical implications will remain valid when extending our approach, in particular regarding the transverse nature of the spin and its relation to the optical responses of the metals at play. While the exact values of $\Omega_\infty$ will be modified at the closest separations between the sphere and the metal film, the discussion will qualitatively remain the same, and the same conclusion holds.

\begin{figure}[htbp]
\centering%\vspace{4cm}
\includegraphics[width=0.42\textwidth]{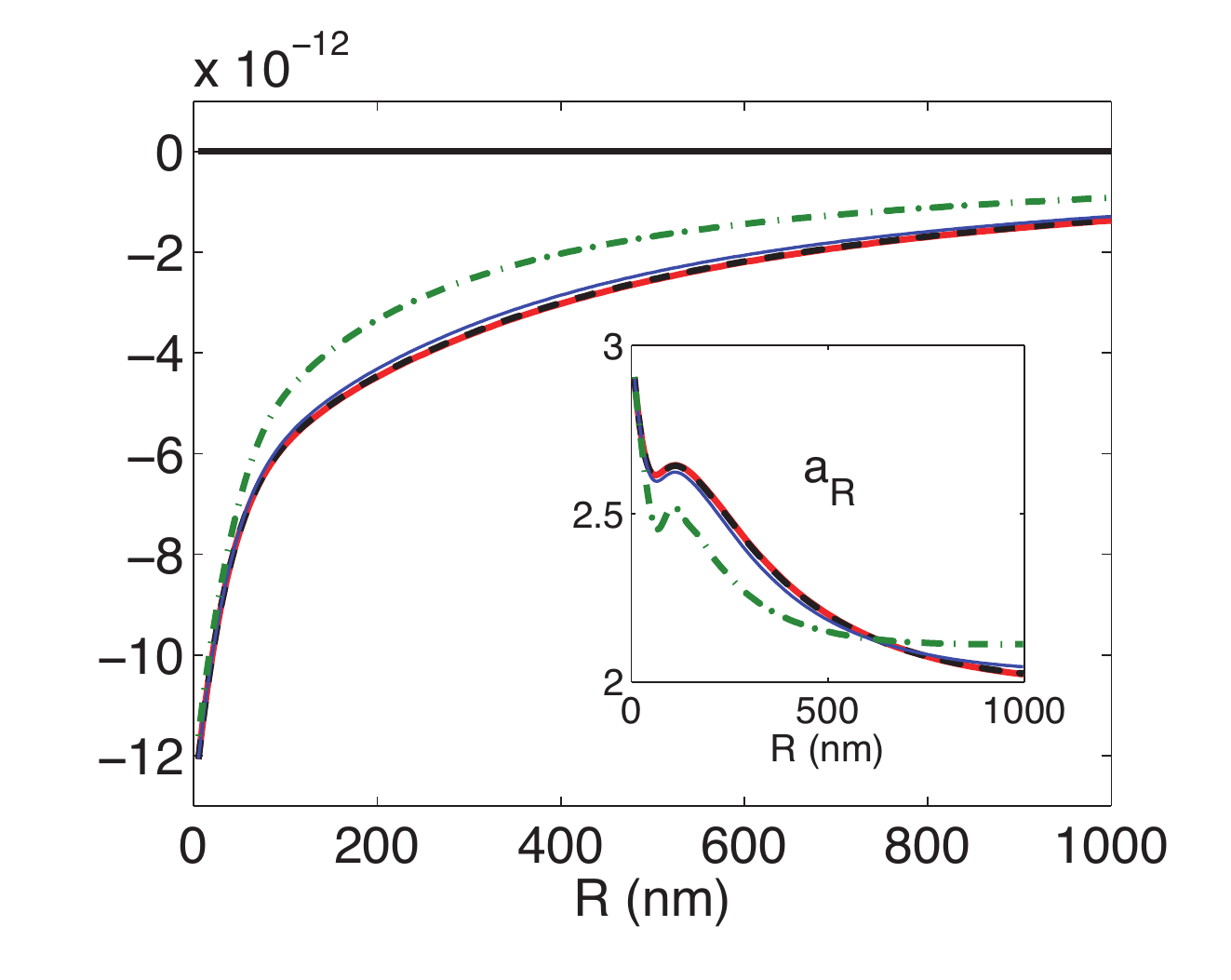} %
\caption{Same as Fig.~\ref{Ny_dielectric}, but the normalized torques are further divided by $R^3$, in order to study the spinning angular velocity $\Omega_\infty$. (insert) Local power law $a_R$ (see Eq.~(\ref{def_alpha})) with respect to the variable $R$ for the torques evaluated for each value of $n''$.}
\label{omega_lim_dielectric}
\end{figure}

The influence of dissipation is studied again here by choosing different values for the dissipation inside dielectric spheres. As seen in Fig.~\ref{omega_lim_dielectric} 
for $n''\neq 0$, the reduced torque $n_y/R^3$ is almost proportional to $n''$, and decreases monotonically when $R$ increases. This clearly means that the torque $N_y$ increases more slowly than $R^3$ and that the spinning angular velocity $\Omega_\infty$ of the sphere will decrease as the sphere gets larger. This dependence is very different from what is obtained for metal spheres and, in order to emphasize this, we introduce the local power law $a_R$
\begin{align}\label{def_alpha}
a_R = \frac{\mathrm{d} \ln n_y}{\mathrm{d} \ln R} ~ ,
\end{align}
which fits locally the curvature of $n_y(R)$ with a power function $R^{a_R}$. This parameter is plotted in the insert of Fig.~\ref{omega_lim_dielectric}, and shows that for small spheres, $a_R$ goes to $3$, in agreement with the dipolar limit given in Eq.~(\ref{Ndip}), for which $\Im[\alpha] \propto R^3$. For larger spheres however, $a_R$ seems to converge to $2$, at least for materials with small dissipation.

\begin{figure}[htbp]
\centering%\vspace{4cm}
\includegraphics[width=0.42\textwidth]{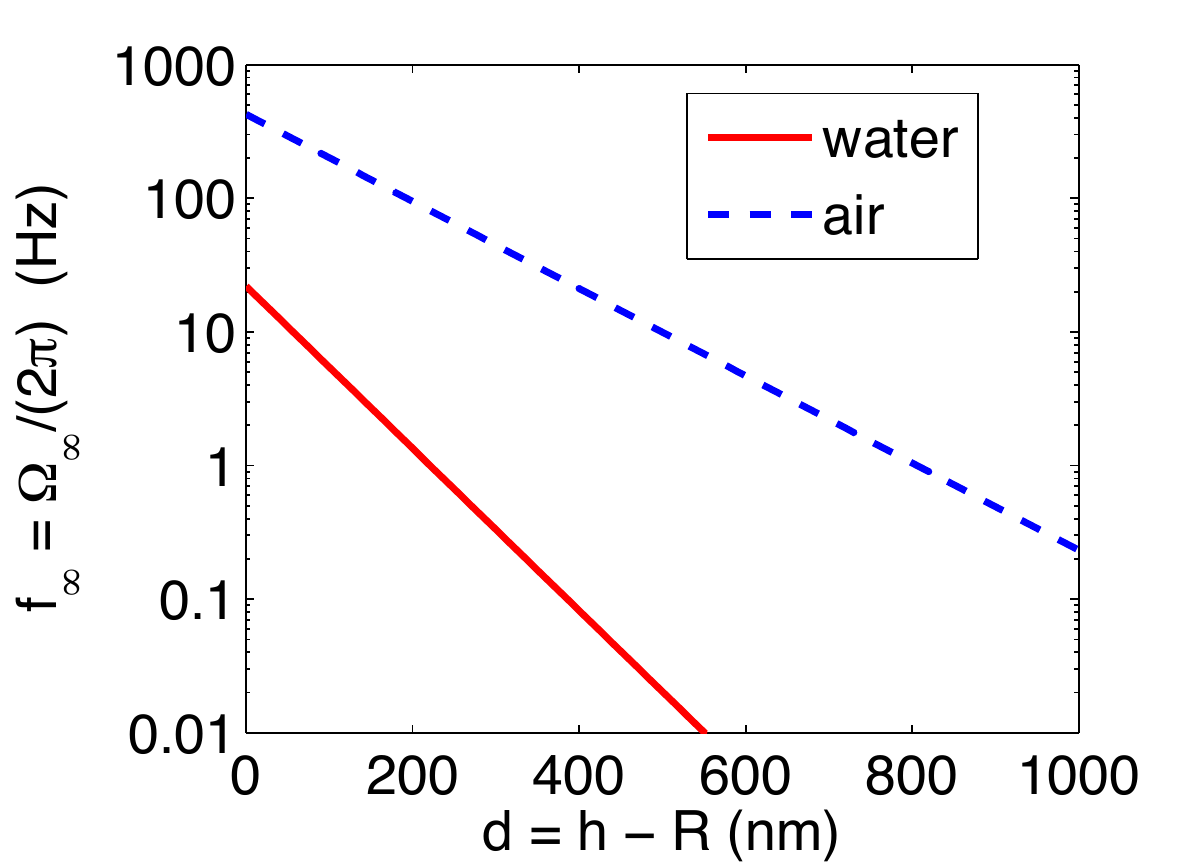} %
\caption{Limiting spinning frequency $f_\infty=\Omega_\infty/2\pi$ of a 500 nm radius gold sphere, as function of the sphere separation to the plasmonic interface $d=h-R$. The surrounding medium is either water or air. The plasmon intensity is $I_0=10^{9}~{\rm W}\cdot {\rm m}^{-2}$.}
\label{fig:flim_waterair}
\end{figure}

An estimation of the spinning frequency $f_\infty = \Omega_\infty /2\pi$ for a 500 nm gold radius sphere is given in Fig.~\ref{fig:flim_waterair} as a function of the distance $d=h-R$ between the sphere and the gold plane.  Two situations are compared, both involving an SP mode excited at $\lambda=594$ nm and a laser intensity of $I_0=10^{9}~{\rm W}\cdot {\rm m}^{-2}$. When the sphere is immersed in water, for which $n_\mathrm{d}=1.33$ and $\eta=10^{-3}$, a rotation frequency $f_\infty \simeq 22$ Hz is reached for a sphere close to contact. When the distance $d$ increases, the amplitude of the torque decreases exponentially as the plasmonic field does away from the interface. The rotation frequency is divided by 10 every 160 nm in the separation distance. When the spinning takes place in air, with  $n_\mathrm{d}=1$ and a much lower viscosity $\eta\simeq 1.8 \cdot 10^{-5}$, inertia is accounted for. There, the rotation speed limit corresponds to much higher $f_\infty$ frequencies, up to $400$ Hz in the vicinity of the plane. Moreover in air, the exponential decay is twice slower when the sphere moves away from the plane.  We finally emphasize that from an experimental point of view, these frequencies perfectly fall within the bandwidths of standard displacement detection systems, based on power spectral analysis or rotational Doppler shifts, for instance \cite{friese1996optical,orrit2011brownian,lavery2013detection}.

\section*{Conclusions}

We evaluate in this article the forces and torques that a surface plasmon can induce on a sphere using a multipolar decomposition of the incident plasmonic field and its reflection on the sphere. While higher order reflections between the plane and the sphere should be included if one aims for precise quantitative evaluations, this first-order approach provides qualitative and meaningful insights in the physics of the problem. Our results show unambiguously the appearance of a plasmonic torque that can spin the sphere. We stress that contrary to usual optical torques, the angular velocity vector induced by the SP mode points in a direction parallel to the plasmonic interface but transverse to the plasmon propagation. For metals, we showed that the torque is expected to be more efficient with gold objects, and at wavelengths where both the spin expectation value of the plasmon and the absorption in the sphere are maximized. This leads us to emphasize the crucial role played by both of them in the spinning process. We finally note that the magnitudes of the induced torques should lead to measurable rotations. These values could be further increased by a more powerful plasmonic field, or by using a less viscous media. %\sout{But before all, it is important to note that the first-order scattering approach most probably underestimates these values and that a more realistic modeling that accounts for multiple reflection effects between the sphere and the metal film will amplify the spin frequency, in particular at the smallest separation distances. Such an approach however lies beyond the scope of this manuscript. }

\section*{Acknowledgements}

We thank T. W. Ebbesen, G. Schnoering and A. Cuche for fruitful discussions. We acknowledge support from the French program Investissement d'Avenir (Equipex Union).

\appendix

\section{The field scattered by the sphere}

\label{app:scattering}

The field $\left( \bcalE^s, \bcalH^s \right)$ scattered by the sphere can be written in the basis of the spherical modes in terms of the Mie coefficients \cite{hulst1957light,bohren2008absorption}. These coefficients can be expressed thanks to ordinary Bessel functions of the first and second kind
\begin{align*} % mêmes conventions que dans thèse, v d Hulst, Bohren-Huffman
&a_\ell(\beta) = \frac{\teps C^{A}_\ell - C^{B}_\ell }{\teps C^{C}_\ell - C^{D}_\ell}
&b_\ell(\beta) = \frac{ C^{A}_\ell - C^{B}_\ell }{ C^{C}_\ell - C^{D}_\ell}
\end{align*} 
where $\teps=\eps / \eps_d$ is the ratio of the permittivities inside and outside the sphere, $\beta=\omega n_d R / c$ is a dimensionless size parameter, and the four coefficients can be expressed thanks to ordinary Bessel functions of the first and second kind
\begin{align*}
&C^{A}_\ell = J_{\ell+1/2}(\tn \beta) \left[ \beta J_{\ell-1/2}(\beta) - \ell J_{\ell+1/2}(\beta) \right] \\
&C^{B}_\ell = J_{\ell+1/2}(\beta) \left[ \tn \beta J_{\ell-1/2}(\tn \beta) - \ell J_{\ell+1/2}(\tn \beta) \right] \\
&C^{C}_\ell = J_{\ell+1/2}(\tn \beta) \left[ \beta H_{\ell-1/2}(\beta) - \ell H_{\ell+1/2}(\beta) \right] \\
&C^{D}_\ell = H_{\ell+1/2}(\beta) \left[ \tn  \beta J_{\ell-1/2}(\tn \beta) - \ell J_{\ell+1/2}(\tn \beta) \right] ~ .
\end{align*}
Using these coefficients, the components of the scattered field can be expressed in a compact form (see for instance \cite{barton1989theoretical} for the complete expressions). We bring the reader's attention to the fact that we have chosen the same convention as in \cite{hulst1957light,bohren2008absorption} for the Mie coefficients, which is opposite in sign with respect to \cite{barton1989theoretical,almaas1995radiation} and does not include the $A_{\ell,m}, B_{\ell,m}$ coefficients, hence the small differences in the expressions of the force and torque compared to the latter articles.

\section{Multipolar decomposition of a plasmonic field}
\label{app:multipolar}

%In order to compute the optical force acting on the nanoparticles beyond the dipolar approximation, we use the scattering formalism to perform a multipolar calculation. In this formalism, the electromagnetic field is described as being made of an incident field $\left( \bE^i, \bH^i \right)$ coming to the sphere, and a scattered field $\left( \bE^s, \bH^s \right)$ going away from it. The total field $\left( \bE, \bH \right)$ outside the sphere is the sum of the two later components. The radiation pressure  exerted on the sphere can then be obtained by integrating over its surface the Maxwell stress-tensor for this total electromagnetic field. 

Because of the spherical symmetry of the system, the fields are expanded in the basis of spherical modes, taking the center of the sphere as the origin of the frame. The calculations detailed here are based on the works \cite{barton1989theoretical,almaas1995radiation}. An alternative method implemented for evanescent fields has been presented in \cite{bekshaev2013mie}.
The incident plasmonic field given in Eq.~(\ref{plasmonic_field2}) can be expressed in terms of spherical modes thanks to the multipolar coefficients $(A_{\ell,m}, B_{\ell,m})$ better suited to describe the scattering on a spherical object. Following the derivations of \cite{barton1989theoretical}, we can obtain these coefficients from the radial components of the fields:
\begin{align}\label{AlmBlm}
&A_{\ell,m} = \frac{(b/R)^2}{\ell (\ell+1) \psi_\ell(k_d b)} \int_\Omega \frac{E^i_r (b,\theta,\phi)}{E_0} Y_{\ell,m}^{*}(\theta,\phi) \sin\theta \d \theta \d \phi \nonumber \\
&B_{\ell,m} = \frac{(b/R)^2}{\ell (\ell+1) \psi_\ell(k_d b)} \int_\Omega \frac{H^i_r (b,\theta,\phi)}{H_0} Y_{\ell,m}^{*}(\theta,\phi) \sin\theta \d \theta \d \phi
\end{align}
where $\psi_\ell$ are the Riccati-Bessel functions, $Y_{\ell,m}$ are the spherical harmonics, $b$ is an arbitrary integration radius, $k_d=\frac{\omega n_d}{c}$ is the wavevector in the surrounding medium,  and $\Omega=[0,\pi]\times[0,2\pi]$. %\footnote{For numerics, $b=\frac{\lambda}{n}$ seems to be a good choice for the sake of stability.}. 
%The Riccati-Bessel function $\psi$ can be expressed easily with the ordinary Bessel function of the first kind $J$, and the spherical harmonics $Y$ with the associated Legendre fonctions $P$:
%%
%\begin{align*}
%&\psi_\ell(k_d b) = \sqrt{\frac{\pi k_d b}{2}} J_{\ell+1/2} (k_d b) \\
%&Y_{\ell,m} (\theta,\phi) = \sqrt{\frac{2\ell+1}{4\pi}} \sqrt{\frac{(\ell-m)!}{(\ell+m)!}} P_\ell^m (\cos\theta) e^{\imath m \phi}~ .
%\end{align*}
%%

In the case of a plasmonic field given in Eq.~(\ref{plasmonic_field2}), similarly to the derivation presented in \cite{almaas1995radiation} for an evanescent wave, it is possible to simplify the calculation of the coefficients $A_{\ell,m},B_{\ell,m}$ in Eq.~(\ref{AlmBlm}) to a 1D-integration. Using spherical coordinates $(r,\theta,\phi)$, with $\theta$ taken with respect to the $z$ axis, and $\phi$ taken with respect to the $x$ axis in the $(x,y)$ plane, we get the following radial components for the incident field on the surface of a sphere with radius~$b$:
\begin{align}\label{plasmonic_radial}
E_r^i (b, \theta,\phi) = & E^i e^{\imath q h} e^{\imath k b \sin\theta \cos\phi} e^{\imath q b \cos\theta} \nonumber \\
& \times \left( \tq \sin\theta \cos\phi - \tk \cos\theta  \right) \nonumber \\
H_r^i (b,\theta,\phi) = & H^i e^{\imath q h} e^{\imath k b \sin\theta \cos\phi} e^{\imath q b \cos\theta} \sin\theta \sin\phi ~ .
 \end{align}
Then, one of the integrations in the determination of the multipolar coefficients can be analytically carried out. By incorporating Eqs.~(\ref{plasmonic_radial}) in the expressions (\ref{AlmBlm}), one can use the following integral representations of the modified Bessel function $I$
\begin{align*}
I_{|m|} (z) = \frac{1}{\pi} \int_0^\pi e^{z\cos\phi} \cos(m\phi) \d \phi ~  , 
\end{align*}
valid for any $z\in \mathds{C}$, to take care of the integration over $\phi$ analytically. We are then left with terms that only involve $\theta$-integrations:
\begin{align}\label{AB_plasmonic}
&A_{\ell,m} = Q_0^{(\ell,m)} e^{\imath q h} \left[ \tq Q_1^{(\ell,m)} - \tk Q_2^{(\ell,m)} \right] \nonumber \\
&B_{\ell,m} = Q_0^{(\ell,m)} e^{\imath q h} Q_3^{(\ell,m)}
\end{align}
with the explicit term
\begin{align}\label{eq_Q0}
Q_0^{(\ell,m)} = \sqrt{\frac{2\ell+1}{4\pi}} \sqrt{\frac{(\ell-m)!}{(\ell+m)!}} \frac{(b/R)^2}{\ell(\ell+1) \psi_\ell(k_d b)}
\end{align}
and the three 1D-integrals:
\begin{align*}%\label{eq_Q123}
Q_1^{(\ell,m)} = &\int_0^\pi \d\theta ~\sin^2 \theta ~e^{\imath q b \cos\theta} P_\ell^m(\cos\theta) \nonumber \\
&\times  \int_0^{2\pi} \d \phi~ e^{\imath k b \sin \theta \cos\phi} \cos\phi ~e^{-\imath m \phi} \nonumber \\
= & 2\pi \int_0^{\pi/2} \d\theta \sin^2\theta \left\{ \begin{array}{c} \cos(q b \cos\theta) \\ \imath \sin(q b \cos\theta)\end{array} \right\}  P_\ell^m (\cos\theta) \nonumber \\
&\times  \left[ I_{|m-1|}(\imath k  b \sin\theta)  +I_{|m+1|}(\imath k  b \sin\theta) \right]  
\end{align*}
\begin{align*}
Q_2^{(\ell,m)} = & \int_0^\pi \d\theta ~\sin\theta \cos\theta ~  e^{\imath q b \cos\theta} P_\ell^m(\cos\theta) \nonumber \\
&\times \int_0^{2\pi} \d \phi~ e^{\imath k b \sin \theta \cos\phi} e^{-\imath m \phi} \nonumber \\
= & 4\pi \int_0^{\pi/2} \d\theta \sin\theta \cos\theta \left\{ \begin{array}{c} \imath \sin(q b \cos\theta) \\ \cos(q b \cos\theta) \end{array} \right\} \nonumber \\
& \times P_\ell^m (\cos\theta)  I_{|m|}(\imath k  b \sin\theta) 
\end{align*}
\begin{align*}
Q_3^{(\ell,m)} =& \int_0^\pi \d\theta ~\sin^2\theta ~e^{\imath q b \cos\theta} P_\ell^m(\cos\theta) \nonumber \\
& \times \int_0^{2\pi} \d \phi~ e^{\imath k b \sin \theta \cos\phi} \sin\phi ~e^{-\imath m \phi} \nonumber \\
=& -4\pi \frac{m}{k  b} \int_0^{\pi/2} \d\theta ~ \sin\theta \left\{ \begin{array}{c} \cos(qb \cos\theta) \\ \imath \sin(q b \cos\theta)\end{array} \right\} \nonumber \\
& \times P_\ell^m (\cos\theta) I_{|m|}(\imath k  b \sin\theta)
\end{align*}
with braces indicating cases where $(\ell+m)$ is $\left\{ \begin{array}{c} \mathrm{even} \\ \mathrm{odd} \end{array} \right\}$.

%
%
% \footnote{Here the convention is opposite in sign with respect to \cite{hulst1957light}.}.
%%
%\begin{align*}
%&a_\ell(x) = -\frac{\tn^2 C^{A}_\ell - C^{B}_\ell }{\tn^2 C^{C}_\ell - C^{D}_\ell}
%&b_\ell(x) = -\frac{C^{A}_\ell - C^{B}_\ell }{C^{C}_\ell - C^{D}_\ell}
%\end{align*} 
%%
%where $\tn=\frac{n_S}{n_d}$ is the ratio of refractive indices inside and outside the sphere, and 
%%
%\begin{align*}
%&C^{A}_\ell = J_{\ell+1/2}(\tn \beta) \left( x J_{\ell-1/2}(\beta) - \ell J_{\ell+1/2}(\beta) \right) \\
%&C^{B}_\ell = J_{\ell+1/2}(\beta) \left( \tn x J_{\ell-1/2}(\tn \beta) - \ell J_{\ell+1/2}(\tn \beta) \right) \\
%&C^{C}_\ell = J_{\ell+1/2}(\tn \beta) \left( x H_{\ell-1/2}(\beta) - \ell H_{\ell+1/2}(\beta) \right) \\
%&C^{D}_\ell = H_{\ell+1/2}(\beta) \left(\tn  x J_{\ell-1/2}(\tn \beta) - \ell J_{\ell+1/2}(\tn \beta) \right)
%\end{align*}
%%
%with $\beta=k_d R$. 

%\section{Derivation of the Brownian noise in rotation}
%\label{section:stochastic}

\bibliographystyle{apsrev}
\bibliography{biblio_mie}

\end{document}